%% file: main.tex
\documentclass[a4paper,11pt]{article}
\usepackage{pos}

\usepackage{physics} % for \bra and \ket
\usepackage{multicol} % for Multicolumn figures
\usepackage{slashed} % for Feynman slash
\usepackage{xcolor,cancel}
\usepackage{bm} % Bold maths for vect
\usepackage{amsmath}
\usepackage{xspace}
\usepackage{hyperref}
\usepackage{todonotes} % for todo boxes
\usepackage{tikz}
\usetikzlibrary{decorations.pathmorphing}
\usetikzlibrary{decorations.markings}

\tikzset{
    photon/.style={thick,black,decorate,decoration={snake,amplitude=2pt,post 
length=0pt,pre
    length=0pt}}, 
photon2/.style={thick,black,decorate,decoration={snake,amplitude=1.3pt,segment 
    length=6.1pt,post length=0pt,pre
    length=0pt}}, 
photon3/.style={thick,black,decorate,decoration={snake,amplitude=1.3pt,segment 
    length=4.95pt,post length=0pt,pre
    length=0pt}}, 
photon4/.style={thick,black,decorate,decoration={snake,amplitude=1.2pt,segment 
    length=5.04pt,post length=0pt,pre
    length=0pt}},     
    quark/.style={thick,black,postaction={decorate},
        decoration={markings,mark=at position .55 with
{\arrow[draw=black]{>}}}},
    thickquark/.style={very thick,black,postaction={decorate},
        decoration={markings,mark=at position .55 with {\arrow[black]{>}}}},
    thickquarkarrow/.style={very thick,black,postaction={decorate},
        decoration={markings,mark=at position .55 with {\arrow[black]{<}}}},
    gluon/.style={thick,decorate, draw=black,
        decoration={coil,amplitude=4pt, segment length=5pt,post length=0pt,pre
    length=2pt}},
    W/.style={thick,black,decorate,decoration={snake,segment length=5pt,post
    length=0pt,pre length=2pt}},
    H/.style={very thick,dashed}, 
}

\def\inf {\ensuremath{\infty}\xspace}
\newcommand{\vect}[1]{\ensuremath{\bm{#1}}}
\newcommand{\conj}[1]{\ensuremath{#1^{\ast}}}

\title{Towards a lattice determination of the form factors of the rare Hyperon decay $\Sigma^+ \to p \ell^+ \ell^-$}
\ShortTitle{$\Sigma^+ \to p \ell^+ \ell^-$ on the lattice}
\author[a]{Felix Erben}
\author[a]{Vera G\"ulpers}
\author*[a]{Raoul Hodgson}
\emailAdd{raoul.hodgson@ed.ac.uk}

\author[a]{Antonin Portelli}
\affiliation[a]{Higgs Centre for Theoretical Physics, University of Edinburgh,\\
  EH9 3FD, Edinburgh, UK}

\abstract{The rare Hyperon decay $\Sigma^+ \to p \ell^+ \ell^-$ is an $s \to d$ flavour changing neutral current, and is therefore suppressed in within the Standard Model, making it an excellent probe for new physics. The process is dominated by long distance processes and therefore lattice QCD is the only existing technique to obtain a first principles Standard Model theoretical prediction. We present our work on an unphysical exploratory study of the rare Hyperon decay on the lattice, including an investigation into the applicability of source-sink sampling for this decay, and we show a generalisation of the summed method for arbitrary 3-point functions and 4-point functions. In addition we show preliminary results for a computation of the $s \to u$ semileptonic Hyperon decays.}

\FullConference{%
 The 38th International Symposium on Lattice Field Theory, LATTICE2021
  26th-30th July, 2021
  Zoom/Gather@Massachusetts Institute of Technology
}

\begin{document}
\maketitle

\section{Introduction}
The rare Hyperon decay $\Sigma^+ \to p \ell^+ \ell^-$ is an $s \to d$ Flavour Changing Neutral Current (FCNC) process that is heavily suppressed within the Standard Model (SM) since FCNCs are forbidden at tree level and therefore must proceed via virtual loops. This makes processes such as the rare Hyperon decay excellent probes of physics beyond the SM (BSM) because they have small SM contributions that could be greatly enhanced by tree level BSM FCNC processes. 

Also of interest are the $s \to u$ quark transitions within the octet baryons ($Y$). Here we focus on the $Y \to Y' \ell^- \bar{\nu}$ charged current semileptonic decays that can proceed at tree level within the SM and are therefore much better suited to precision measurements. Combining experimental measurements with theoretical predictions allows for the measurement  of the CKM matrix element $V_{us}$. This provides an alternative determination from the $K_{\ell3}$ and $K_{\ell2}$ decays.

\subsection{Experimental Motivation}
The first experimental evidence for the $\Sigma^+ \to p \mu^+ \mu^-$ decay was provided by the HyperCP experiment \cite{HyperCPCollaboration2005Evidencemu} at Fermilab in 2005 in which they observed three events and measured the branching fraction to be $\mathcal{B}(\Sigma^+ \to p \mu^+ \mu^-) = 8.6^{+6.6\ +5.5}_{-5.4\ -5.5} \times 10^{-8}$. This value is consistent with the SM prediction (see section \ref{Sec:Theo_Motivation}), however the three events were very close in dimuon invariant mass. HyperCP claimed this as potential evidence for the existence of a new intermediate particle of mass $214.3(4) MeV$, and therefore new physics. This became known as the HyperCP anomaly.

In 2018, the LHCb collaboration made their first observation of the same decay with approximately 10 candidate events. From this they measured the branching fraction to be $\mathcal{B}(\Sigma^+ \to p \mu^+ \mu^-) = 2.2^{+0.9\ +1.5}_{-0.8 \ -1.1} \times 10^{-8}$ which is consistent with both the HyperCP measurements and the SM prediction. In addition, they were able to measure the dimuon invariant mass distribution with greater accuracy than the HyperCP experiment and found no evidence of the resonance peak seen previously, contradicting the claim of new physics. 
        
\subsection{Theory Motivation}
\label{Sec:Theo_Motivation}
Having a theoretical SM prediction of this decay is of vital importance when searching for BSM physics. The theoretical prediction of the branching fraction was first calculated in 2005 \cite{He2005DecayModel} and updated in 2018 \cite{He2018DecayPmu+mu-} (with negligible numerical change). Their result is given by the range
\begin{eqnarray}
\label{eqn:SM_bounds1}
    1.6 \times 10^{-8} \leq \mathcal{B}(\Sigma^+ \to p \mu^+ \mu^-) \leq 9.0 \times 10^{-8}.
\end{eqnarray}
It was found that the short distance contribution to the branching ratio is $\mathcal{B}_{SD} \sim 10^{-12}$ which is negligible compared to the long distance contribution. This long distance contribution was computed using a combination of ChPT, unitarity cuts, experimental inputs and vector meson dominance models. The large range of values in equation \eqref{eqn:SM_bounds1} stems from the fact that different approaches in ChPT give differing results and therefore a precise prediction cannot be made with these methods.

Since the first observation of the muonic decay mode, the LHCb experiment has planned improvements \cite{AlvesJunior2019ProspectsLHCb}, such as a dedicated trigger for this decay, that will allow for significantly improved precision as well as first measurements of asymmetry observables, differential decay rates and the $\Sigma^+ \to p e^+ e^-$ channel. It is therefore of importance to the experimental community that the theoretical uncertainties in this decay are reduced in order to draw new conclusions once this new data is obtained. Lattice QCD is currently the only tool available for making a model independent ab initio prediction of the SM contribution that can allow for the currently dominant uncertainty to be removed.

\section{Lattice Theory}
The long distance part of the rare Hyperon decay is dominated by the intermediate virtual photon transition $\Sigma^+ \to p \gamma^{\ast}$, the hadronic part of which is given by the amplitude
\begin{align}
    \mathcal{A}_{\mu}^{rs} = \int d^3 \vect{x} \bra{N,\vect{p}_N,r} T \left\{ H_w(x) J_{\mu}(0) \right\} \ket{\Sigma,\vect{p_\Sigma},s}
\end{align}
where $N$ is the nucleon, $s$ and $r$ are the spin component of the baryons, $\vect{p}_{\Sigma,N}$ are the two momenta, $J_{\mu}$ is the electromagnetic (e.m.) current, and $H_w$ is the $s \to d$ effective weak Hamiltonian given by the 4-quark operators $Q^q_{1,2}$ \cite{Buchalla1995WeakLogarithms}
\begin{align}
    H_w(x) = \frac{G_F}{\sqrt{2}} V_{us}^{\ast} V_{ud} \left[ C_1(Q_1^u - Q_1^c) + C_2(Q_2^u - Q_2^c) + ...\right] + \text{c.c.} \\
     Q_1^q = (\bar{s}\gamma_{\mu}^L d) (\bar{q} \gamma_{\mu}^L q) \hspace{2em} Q_2^q = (\bar{s}\gamma_{\mu}^L q) (\bar{q} \gamma_{\mu}^L d). \hspace{4em}
\end{align}
This transition amplitude can be written in terms of four form factors $f_{1,2}$ and $ g_{1,2}$ as such
\begin{align}
    \mathcal{A}_{\mu}^{rs} & = \bar{u}_N^{r}(p_N) \left[ \left(q^2 \gamma_{\mu}-\slashed{q} q_{\mu} \right) (f_1+g_1\gamma_5) + \sigma_{\mu \nu} q^{\nu} (f_2+g_2\gamma_5) \right] u_{\Sigma}^{s}(p_\Sigma)
\end{align}
where $q=p_{\Sigma}-p_N$ is the 4-momentum transfer of the photon and $u_{\Sigma,N}$ are the spinors associated with the $\Sigma$ and $N$ particles respectively.
On the lattice, these form factors can be extracted from the 4-point correlation function
\begin{align}
    C^{(4)}_{\mu}(\Delta t, t_H, t_J, \vect{p}_\Sigma, \vect{p}_N) & = \int d^3 \vect{y} \int d^3 \vect{x} e^{-i \vect{q}\cdot \vect{x}}\langle \psi_N(\Delta t, \vect{p}_N) H_w(t_H,\vect{y}) J_{\mu}(t_J,\vect{x}) \bar{\psi}_\Sigma(0,\vect{p}_\Sigma) \rangle
\end{align}
where $\psi_{\Sigma,N}(t,\vect{p})$ are interpolators for the $\Sigma$ and $N$ states located at time $t$ with spatial momentum $\vect{p}$, and $\Delta t$ is the temporal separation between the source and sink.
This correlator is computed with four types of weak Hamiltonian contractions, as shown in figure \ref{Fig:S2N_Hw_3pt}, with an additional e.m. current insertion on each of the quark legs and a quark disconnected contribution.
\input{all_3pt}

In order to perform this calculation, we use the open source C++ libraries Grid \cite{Boyle2016Grid} and Hadrons \cite{Portelli2020Hadrons}, where all of the relevant contractions have been fully implemented.

\subsection{Computational strategy}
\label{Sec:RareK_approach}
Since the rare Hyperon decay is a baryonic equivalent to the rare Kaon decay \cite{Christ2015Longell,Christ2015ProspectsDecays,Christ2016FirstKtopiell+ell-,Christ2016Progressell-}, its computation can proceed via similar methods. This involves fixing the source-sink separation $\Delta t$ and the e.m. current at time $t_J$, and integrating the weak Hamiltonian in a window around this current
\begin{align}
    I_{\mu}^{(4)}(T_a,T_b,\vect{p}_\Sigma, \vect{p}_N) = e^{-(E_N(\vect{p}_N) - E_\Sigma(\vect{p}_\Sigma))t_J} \int_{t_J-T_a}^{t_J+T_b} dt_H \widetilde{C}^{(4)}_{\mu}(t_H, t_J, \vect{p}_\Sigma, \vect{p}_N) \\
    \widetilde{C}^{(4)}_{\mu}(t_H, t_J, \vect{p}_\Sigma, \vect{p}_N) = \frac{C^{(4)}_{\mu}(\Delta t,t_H, t_J, \vect{p}_\Sigma, \vect{p}_N)}{Z_N \conj{Z}_\Sigma e^{-E_N(\vect{p_N}) \Delta t}}. \hspace{6em} \label{Eqn:C4_tilde}
\end{align}
where $Z_X u_X^s= \bra{0} \psi_X \ket{X,s}$ is the interpolator overlap factor for the $X=N \ \text{or} \ \Sigma$ baryon with spin polarisation $s$.
Equation \eqref{Eqn:C4_tilde} removes these overlap factors as well as the $\Delta t$ dependence so long as the source-sink separation is large enough for the excited external states to decay away. This then gives the spectral decomposition of the integrated 4-point function
\begin{align}
    I_{\mu \ r,s}^{(4)}(T_a,T_b,\vect{p}_\Sigma, \vect{p}_N) = & - \sum_n \frac{1}{2E_n} u_N^r(\vect{p}_N) \frac{\bra{N,\vect{p}_N,r} J_{\mu} \ket{n,\vect{p}_\Sigma} \bra{n,\vect{p}_\Sigma} H_w \ket{\Sigma,\vect{p}_\Sigma,s} }{E_\Sigma(\vect{p}_\Sigma)-E_n} \bar{u}^s_\Sigma(\vect{p}_\Sigma) \nonumber \\
    & \hspace{16em} \times (1-e^{-(E_n - E_\Sigma(\vect{p}_\Sigma)) T_a}) \nonumber \\
    & + \sum_m \frac{1}{2E_m} u_N^r(\vect{p}_N) \frac{\bra{N,\vect{p}_N,r} H_w \ket{m,\vect{p}_N} \bra{m,\vect{p}_N} J_{\mu} \ket{\Sigma,\vect{p}_\Sigma,s} }{E_m - E_N(\vect{p}_N)} \bar{u}^s_\Sigma(\vect{p}_\Sigma) \nonumber \\
    & \hspace{16em} \times (1-e^{-(E_m - E_N(\vect{p}_N)) T_b})
    \label{Eqn:Integrated_4pt}
\end{align}
where the sum over intermediate states $m$ contains an implicit sum over spin polarisation.
In order to extract the amplitude, we take the limit $T_a,T_b \to \inf$ so that the additional exponential terms tend to zero. However, there exist intermediate states where $E_n < E_\Sigma(\vect{p}_\Sigma)$, and therefore these terms grow exponentially in $T_a$. For the rare Hyperon decay at the physical point, these states are the $N$ and $N \pi$ states which must be removed from equation \eqref{Eqn:Integrated_4pt} before taking $T_a \to \inf$. The $N$ intermediate state can be removed explicitly by combining matrix elements
\begin{align}
    \bra{N(\vect{p}_N),r} J_{\mu} \ket{N(\vect{p}_\Sigma),s} \text{ and } \bra{N(\vect{p}_\Sigma),r} H_w \ket{\Sigma(\vect{p}_\Sigma),s}
\end{align}
which can be obtained from 3-point functions with insertions of the e.m. current and weak Hamiltonian respectively.

The $N\pi$ state can then be constructed in a similar manner to the $\pi \pi$ state removal in \cite{Christ2019KNuNu}. In addition, there are finite volume corrections that stem from these same $N \pi$ intermediate state. Therefore a study into the $\Sigma \to N \pi$ and $N \pi \to N$ transitions are required to account for these effects at the physical point.

\section{Summed Method}
\subsection{3-point summed method}
The so-called summed method is a technique in which the operator insertion in a 3-point function $\langle \psi(\Delta t) \mathcal{O}(t) \bar{\psi}(0) \rangle$ is summed over the temporal extent of the lattice \cite{Maiani:1987by,Bouchard:2016heu}. For simple insertions of di-quark operators ($\bar{q}\Gamma q$), this can be efficiently achieved via a sequential solve of the propagators involved. 
The summed method has the advantage that excited state contributions to the correlator are exponentially suppressed in $\Delta t$ rather than in $t$ and $\Delta t-t$ as in the fixed t method. In addition, it is significantly computationally cheaper than fixing $t$ in the sequential solve and having to re-invert the Dirac operator for every choice of $t$.
These are both qualities particularly beneficial to baryonic correlators since they suffer from both exponentially falling signal-to-noise ratio, and large excited state contamination. This means there is generally only a very small window in source-sink separations for which a reliable and precise measurement can be made.

However, so far this method has only been used in situations where the initial and final states have identical energies. That is, either the initial and final states are identical, or they are degenerate (e.g. $p$ and $n$ in iso-symmetric limit with momenta such that $|\vect{p}_f| = |\vect{p}_i|$). 
However, one can in theory use this method for arbitrary initial and final states. In this case the integrated 3-point function becomes \cite{Bouchard:2016heu}
\begin{align}
    I^{(3)}(\Delta t) & = \sum_{t=0}^{T} \  \langle \psi_f(\Delta t) J(t) \bar{\psi}_i(0) \rangle \\
            & = \sum_{n,m} \left[ Z^i_{0n} Z^f_{m0} J_{nm} \frac{1-e^{-(E_n-E_m)(\Delta t-1)}}{e^{E_n-E_m}-1} e^{-E_m \Delta t} \right] + \sum_k Z^i_{0k} Z^f_{k0} J_{00} e^{-E_k \Delta t} \\
            & + \sum_n \left[ Z^i_{0n}Z^{fJ}_{n0} + \sum_l \frac{Z^i_{0n} Z^f_{nl}J_{l0}}{e^{E_l}-1} \right]e^{-E_n \Delta t} + \sum_m \left[ Z^{iJ}_{0m} Z^f_{m0} + \sum_l \frac{J_{0l} Z^i_{lm} Z^f_{m0}}{e^{E_l}-1} \right] e^{-E_m \Delta t} \nonumber
\end{align}
where $Z^X_{nm}=\bra{n} \psi_X \ket{m}$, $Z^{XJ}_{nm}=\bra{n} \psi_X J \ket{m}$, $J_{nm} = \bra{n} J \ket{m}$ and the state $\ket{0}$ represents the vacuum.
The third and fourth terms come from contact terms where $t=0\text{ or } \Delta t$ and out of order terms where $t>\Delta t$. Note that there is a $\bra{0} J \ket{0}$ vacuum term what is only present for initial and final states that share identical quantum numbers.

When taken to large $\Delta t$ to allow excited states to decay away, the first term leaves the matrix element of interest up to the overlap factors $Z$, and the exponential time dependence modulated by a function of the energy difference
\begin{align}
    \chi(\Delta,t) = \frac{1-e^{-\Delta(t-1)}}{e^{\Delta}-1}
\end{align}
In the limit where the initial and final states are degenerate $\Delta \to 0$, this simply becomes $\chi(0,t) = t-1$ which is the well known linear envelope used in the summed method to date. However, the full form can in principle be used in the case of flavour changing operators and/or arbitrary momentum transfer.

In addition we are working towards generalising the summed method for use on 4-point functions which could benefit both the rare Hyperon and rare Kaon decay calculations.

\section{Rare Hyperon preliminary results}
We are currently working towards a first exploratory computation of the rare Hyperon decay on an ensemble with a heavier than physical Pion mass. This is the RBC-UKQCD 2+1 flavour $24^3\times 64$ Iwasaki gauge configuration with inverse lattice spacing $a^{-1} = 1780 MeV$ and $m_{\pi} \simeq 340 MeV$ using Shamir domain wall fermions \cite{RBC:2010qam}. 
% The use of chiral fermions prevents mixing of the left-handed 4-quark operators with the right handed operators during renormalisation. 

\begin{figure}[ht]
    \centering
    \begin{multicols}{2}
		\includegraphics[width=0.45\textwidth]{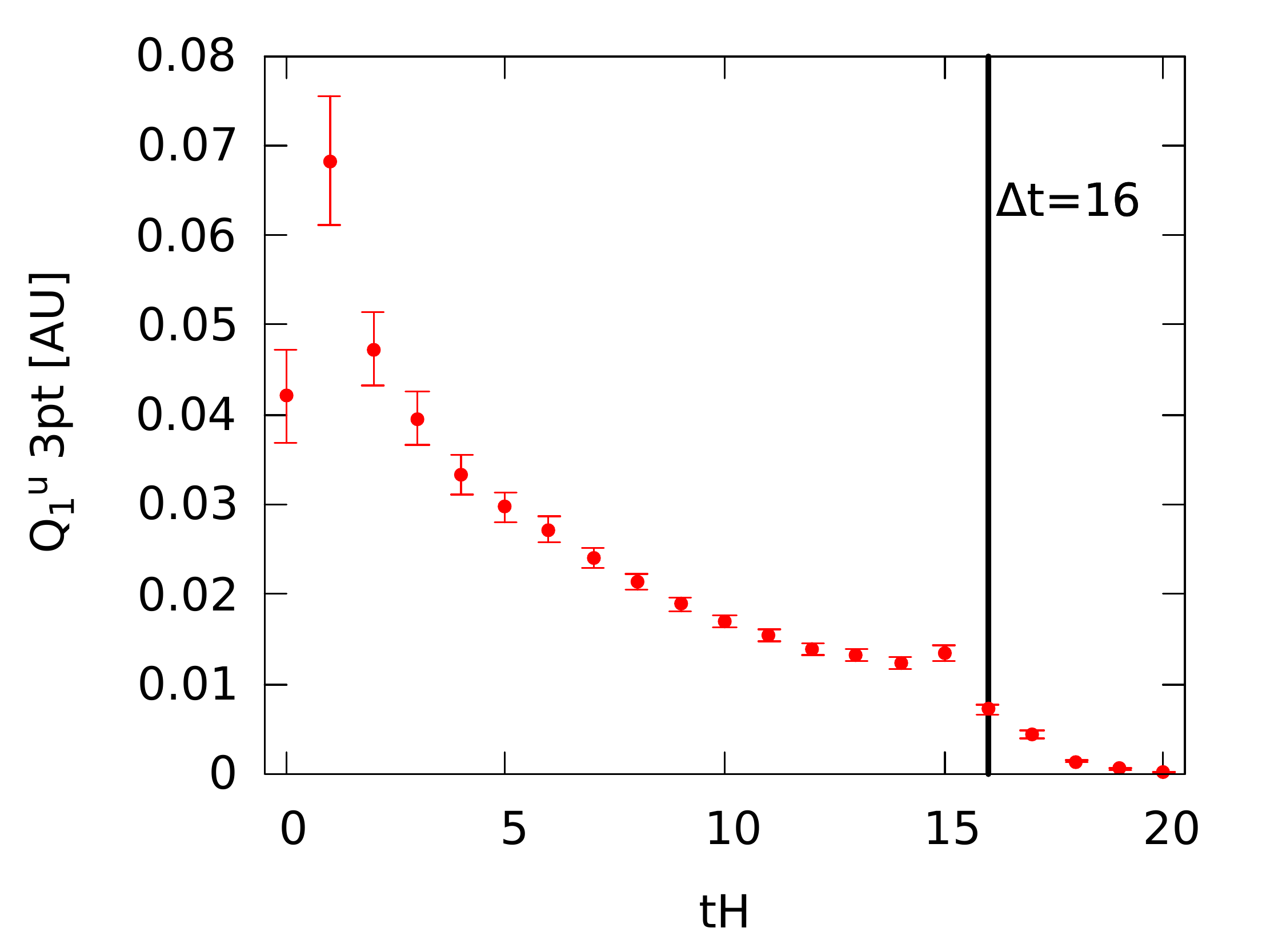}\par
		\includegraphics[width=0.45\textwidth]{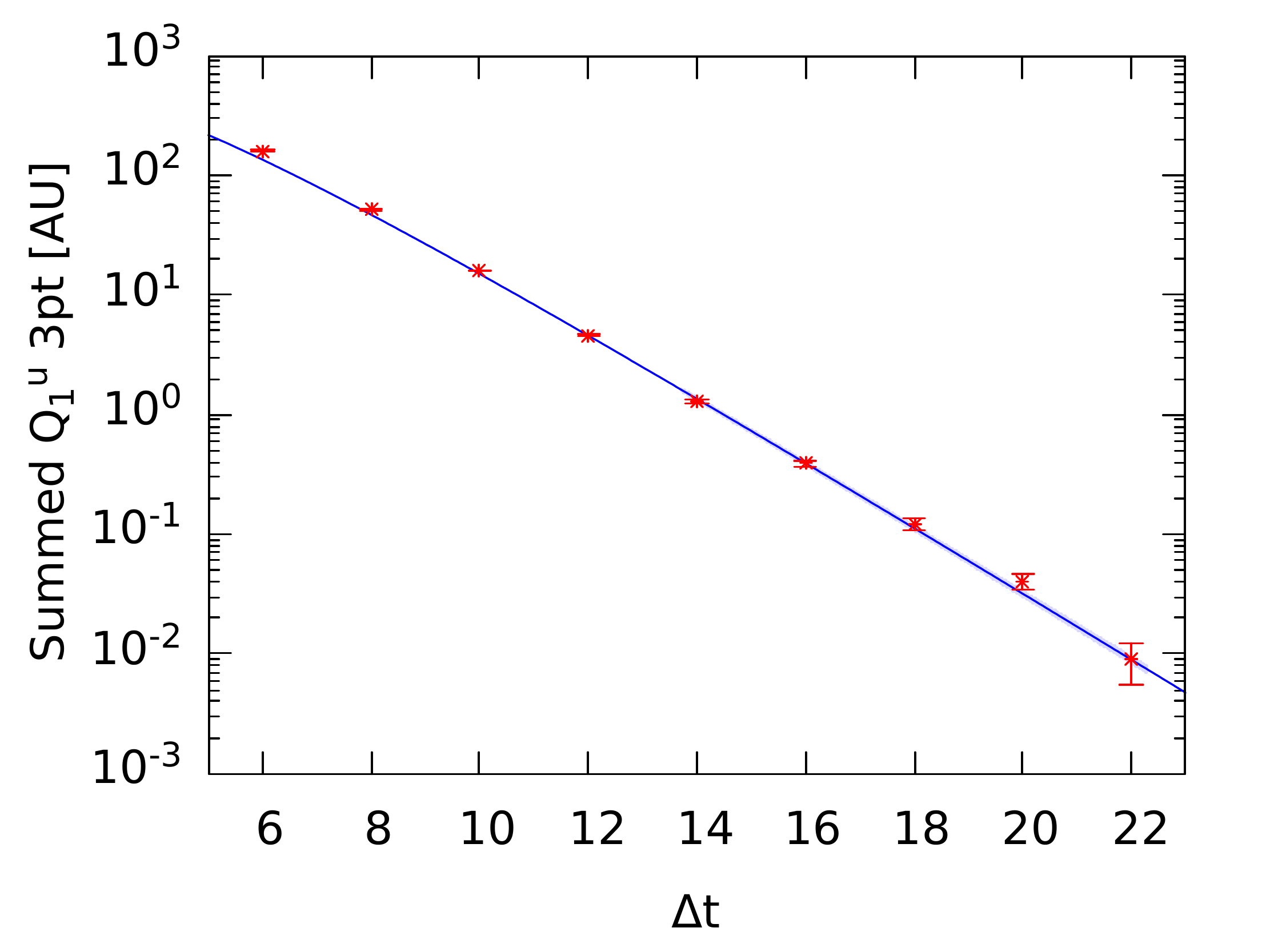}
	\end{multicols}
	\caption{3-point correlator of the connected contribution to $Q_1^u$ with a fixed $\Delta t =16$ (left) and using the summed method (right). Results are shown in arbitrary units.}
	\label{Fig:s2d_corrs}
\end{figure}

Presently we have computed the $C_{sd}$ and $C_{su}$ type diagrams for the 3-point functions of the weak Hamiltonian with both the $\Sigma$ and $N$ at rest. Figure \ref{Fig:s2d_corrs} shows the correlator for the $C_{sd}$ contribution to the 4-quark operator $Q_1^u$. Here we present both the unsummed correlator (left) where the source-sink separation is fixed at $\Delta t=16$, as well as using the summed method (right) leaving $\Delta t$ as the free parameter.
The next step in the calculation is to compute the eye and saucer diagrams of the 3-point function as well as adding the e.m. current insertions on each of the legs.

For this exploratory study we don't need to consider the $N \pi$ intermediate states since these lie above the $E_\Sigma$ threshold on this unphysical ensemble, and therefore decays exponentially in $T_a$ leaving only the growing $N$ intermediate state to be removed.

\section{Source-sink sampling}
In performing the computation of the rare Hyperon diagrams, the position of both the source and the sink are fixed in the propagator solves. Therefore, in order to perform an exact volume sum for a definite momentum projection, this would require $\sim 14,000$ solves on our $24^3\times 64$ lattice. In addition the sum over final state position reduces the error by combining many correlated measurements of the correlator.

\begin{figure}[h]
    \centering
    \begin{multicols}{2}
        \includegraphics[width=0.45\textwidth]{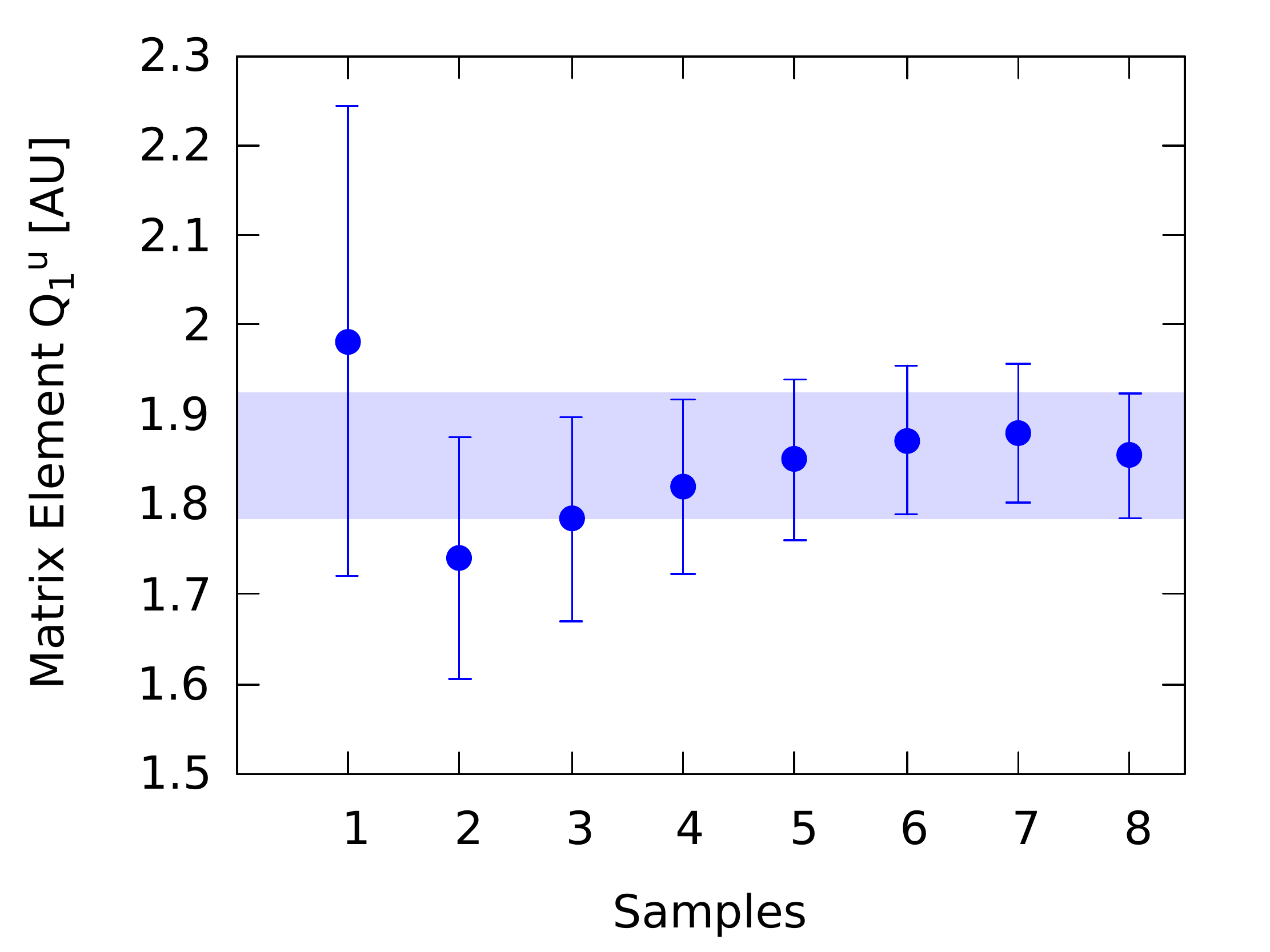}\par
        \includegraphics[width=0.45\textwidth]{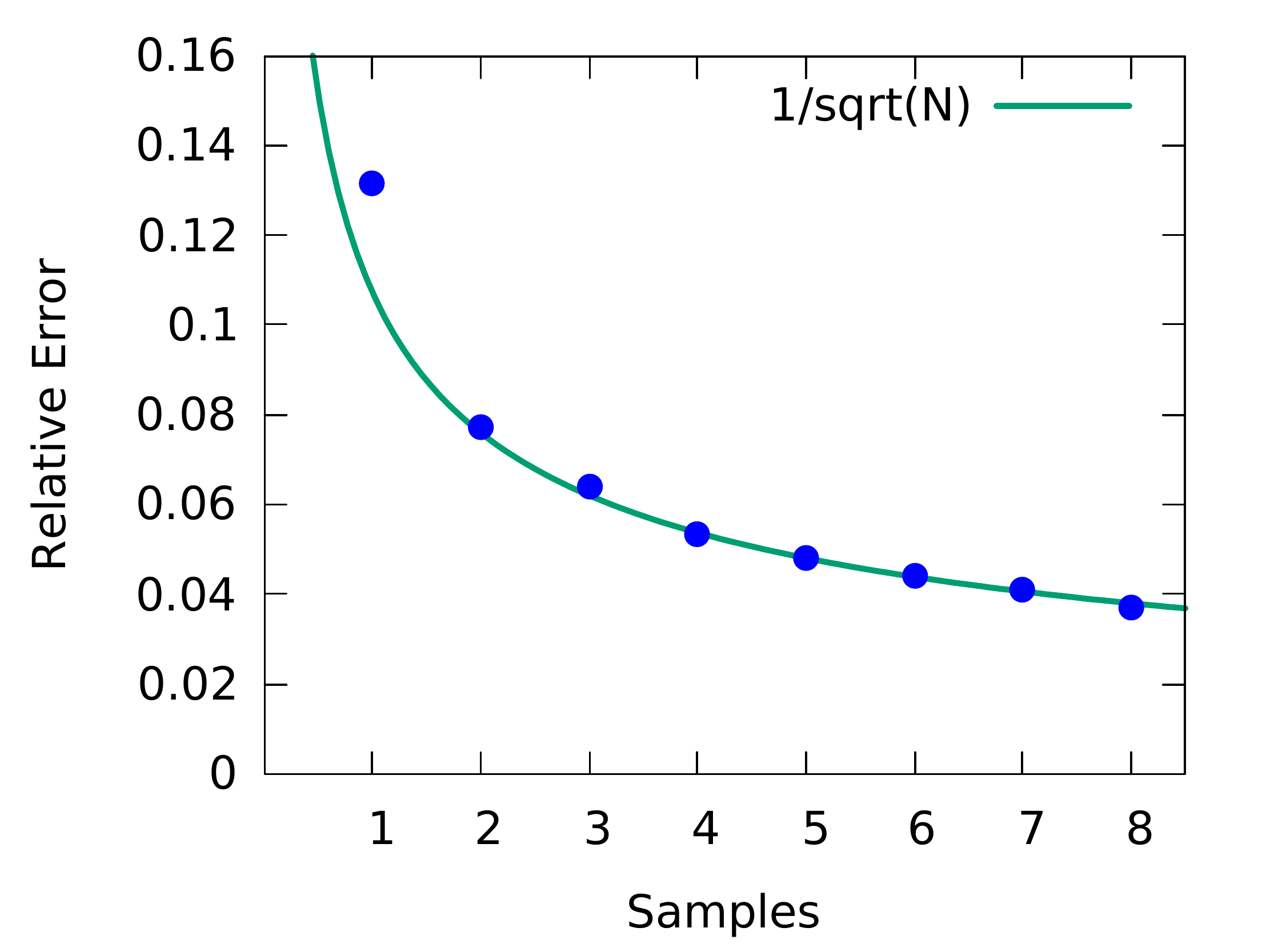}
    \end{multicols}
    \caption{$Q^u_1$ matrix element value (left) and relative error (right) scaling with the number of samples using for source-sink sampling.}
    \label{Fig:S2N_Sampling}
\end{figure}

We can instead approximate this full volume by a sum over $N_S$ randomly selected lattice sites. 
Defining $\Lambda$ to be the set of spatial lattice sites with extent $L$, we can select a subset $\Lambda_S$ that consists only of the randomly selected sites. Therefore we can make the approximation of the sum of some arbitrary function $f(\vect{x})$ \cite{Li:2020hbj}
\begin{align}
    \sum_{\vect{x} \in \Lambda} f(\vect{x}) \simeq \frac{L^3}{N_S} \sum_{\vect{x} \in \Lambda_S} f(\vect{x})
\end{align}
In practice, we can improve this sampling by selecting an independent random set on each timeslice of the lattice $\Lambda_S(t)$.
Assuming a perfect $\mathcal{O}(1/\sqrt{N})$ scaling in the error, performing this sampling at both the source and the sink could achieve up to $\order{1/N}$ scaling due to the independent $\Lambda_S$ at temporally separated points.
However, for the weak Hamiltonian matrix element, we only observe approximately $\mathcal{O}(1/\sqrt{N})$ scaling as can be seen in figure \ref{Fig:S2N_Sampling} (right). This is likely due to large correlations between the samples.

In addition, due to the approximation of the sum, an exact momentum projection cannot be made. Therefore it is possible for additional momentum modes to contaminate the correlator and therefore alter the value extracted for the matrix element.
From figure \ref{Fig:S2N_Sampling} (left), we don't see any statistically significant contamination from other momentum modes above the statistical uncertainty of the measurement.

\section{$s \to u$ Semileptonics}
In addition to the rare Hyperon decay, we have an additional project to compute the $SU(3)$ flavour breaking corrections to the $s \to u$ quark transition $Y \to Y' \ell^- \bar{\nu}$ semileptonic Hyperon decays. We are preforming this calculation at close to physical Pion mass on RBC-UKQCD's $2+1$ flavour $48^3\times 96$ Iwasaki gauge configurations with inverse lattice spacing $a^{-1} = 1730 MeV$ and $m_{\pi} \simeq 140 MeV$ using M\"obius domain wall fermions \cite{RBC:2014ntl}. 

The hadronic part of the matrix element of these decays can be parameterised by six form-factors $f_{1,2,3}$ and $g_{1,2,3}$
\begin{align}
    \bra{Y',\vect{p}',r} J_{\mu} \ket{Y,\vect{p},s} = & \ \bar{u}^r_{Y'}(\vect{p}') \left[ f_1(q^2) \gamma_{\mu} -i f_2(q^2) \frac{ \sigma_{\mu \nu} q^{\nu}}{M_Y+M_{Y'}} + f_3(q^2) \frac{q_{\mu}}{M_Y+M_{Y'}} \right. \\
    & \hspace{2em} \left. + g_1(q^2) \gamma_{\mu} \gamma_5 -i g_2(q^2) \frac{\sigma_{\mu \nu} q^{\nu}}{M_Y+M_{Y'}}\gamma_5 + g_3(q^2) \frac{q_{\mu}}{M_Y+M_{Y'}} \gamma_5 \right] u^s_{Y}(\vect{p}) \nonumber
\end{align}
where $J_{\mu} = \bar{u} \gamma_\mu(1-\gamma_5) s$ is the V-A $s \to u$ quark current, $q=p_Y-p_{Y'}$ is the momentum transfer, and $M_Y$ and $M_{Y'}$ are the masses of the $Y$ and $Y'$ baryons respectively.
These forms factors are known analytically at $q^2=0$ in the $SU(3)$ limit \cite{Cabibbo2003SEMILEPTONICDECAYS}, however, in order to move towards precision measurements of $V_{us}$ from Hyperon decays, we require knowledge of these form factors away from the $SU(3)$ limit.

\begin{figure}[h]
    \centering
    \includegraphics[width=0.6\textwidth]{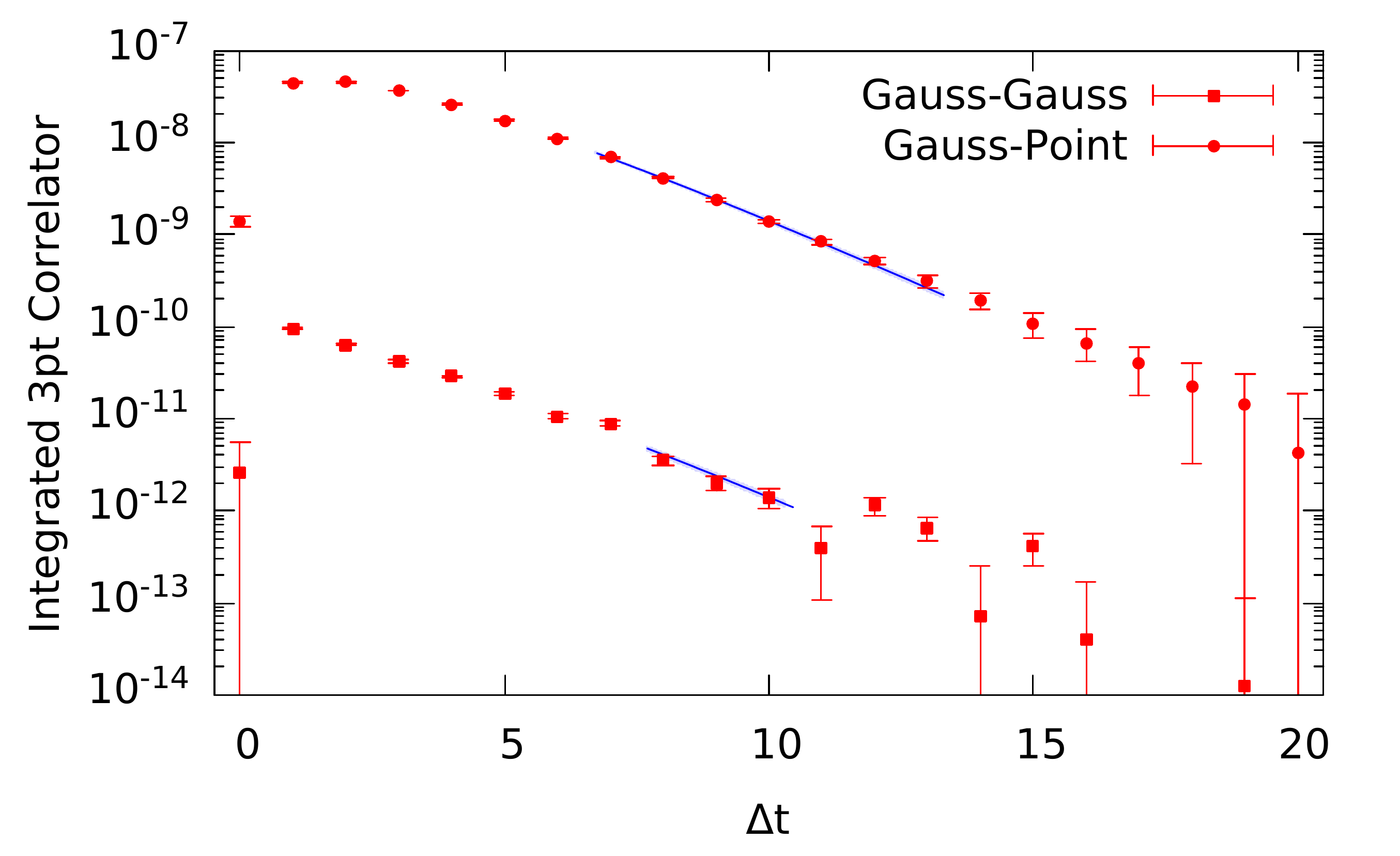}
    \caption{Preliminary example correlator of the $\Sigma^- \to n \ell \bar{\nu}$ decay using the 3-point summed method.}
    \label{Fig:s2u_3pt}
\end{figure}

This project aims to compute the form factors for all of the octet baryon $s \to u$ transitions at the $q^2 = 0$ kinematic point. We are using the summed method with arbitrary external states to reduce the cost of this computation. 
Figure \ref{Fig:s2u_3pt} shows a preliminary example of the correlators for the $\Sigma^- \to n \ell \bar{\nu}$ decay with Gaussian smeared sources as well as Gaussian and point sinks.

\section{Conclusions and Outlook}
We are working towards a first exploratory calculation of the SM contribution to the rare Hyperon decay $\Sigma^+ \to p \ell^+ \ell^-$. This being done on an unphysical ensemble in preparation for a later physical point computation.
It is possible to use similar methods to those used in the computation of the rare Kaon decay \cite{Christ2015Longell,Christ2015ProspectsDecays,Christ2016FirstKtopiell+ell-,Christ2016Progressell-}, however there are additional techniques that could be used to improve the signal and reduce the cost of the computation.
One potential technique is the use of the summed method generalised to two distinct operator insertions and arbitrary initial and final states.
In addition we can use source-sink sampling on the correlation functions to improve the signal of our extracted matrix elements and perform a momentum projection without having to take the prohibitive cost of solving propagators at all spatial points of the lattice. 

In addition, we are working towards a physical point computation of the $SU(3)$ flavour breaking corrections to the flavour changing $s \to u$ semileptonic Hyperon decays. These can produce an alternative measurement of the CKM matrix element $V_{us}$. The computation is being performed using the generalised 3-point summed method where the current is summed over the whole lattice.

% \newpage
% \appendix 

\acknowledgments
% \todo[inline]{What to put here?}
% \noindent The authors thank the members of the RBC and UKQCD Collaborations ...

% \noindent R.H received funding from the European Research Council (ERC) under the European Union’s Horizon 2020 research and innovation programme under grant agreement No 757646.

The authors thank the members of the RBC and UKQCD Collaborations for helpful discussions and suggestions. 
This work used the DiRAC Extreme Scaling service at the University of Edinburgh, operated by the Edinburgh Parallel Computing Centre on behalf of the STFC DiRAC HPC Facility (\href{www.dirac.ac.uk}{www.dirac.ac.uk}). This equipment was funded by BEIS capital funding via STFC capital grant ST/R00238X/1 and STFC DiRAC Operations grant ST/R001006/1. DiRAC is part of the National e-Infrastructure. 
F.E., V.G. and A.P. are supported in part by UK STFC grant ST/P000630/1. 
F.E., V.G., R.H and A.P. also received funding from the European Research Council (ERC) under the European Union’s Horizon 2020 research and innovation programme under grant agreements No 757646 \& A.P. additionally by grant agreement 813942.

% R.H., A.P. and F.E. received funding from the European Research Council (ERC) under the European Union’s Horizon 2020 research and innovation programme under grant agreement No 757646 \& A.P. additionally by grant agreement 813942.

\bibliographystyle{JHEP}
\bibliography{References}

\end{document}

%% file: all_3pt.tex
\begin{figure}
    \centering
    
 \begin{tikzpicture}[scale=0.67]
 %
 %connected
 %
 \draw[bend right=45] (2,0.28) to (0,0);
  \draw[bend right=45] (4,0) to (2,0.28);
   \draw[bend left=35] (2,0.12) to (0,0);
  \draw[bend left=35] (4,0) to (2,0.12);
 %\draw[thick] (4,0) to (0,0);
 \draw[bend right=45] (0,0) to (4,0);
\fill (0,0) circle (0.06cm);
\fill (4,0) circle (0.06cm);
%Weak Hamiltonian
\fill (2,0.12) circle (0.06cm);
\fill (2,0.28) circle (0.06cm);
 \node at (-0.3,-0.3) {\scriptsize $\Sigma^+$};
  \node at (4.2,-0.3) {\scriptsize $P$};
  \node at (2,0.8) {\scriptsize $H_W$};
% quarks
 \node at (1.0,0.38) {\tiny $s$};
  \node at (3.0,0.38) {\tiny $d$};
  \node at (3.0,-0.17) {\tiny $u$};
 \node at (1.0,-0.17) {\tiny $u$};
  \node at (2.0,-0.68) {\tiny $u$};
  \node at (2.0,-1.5) {\small C$_{sd}$};
%
%crossed connected
%
\begin{scope}[xshift=5.5cm]
  \draw[bend right=48] (2.1,0.2) to (0,0);
  \draw[bend right=48] (4,0) to (1.9,0.2);
   \draw[bend left=45] (1.9,0.2) to (0,0);
  \draw[bend left=45] (4,0) to (2.1,0.2);
 %\draw[thick] (4,0) to (0,0);
 \draw[bend right=45] (0,0) to (4,0);
\fill (0,0) circle (0.06cm);
\fill (4,0) circle (0.06cm);
%Weak Hamiltonian
\fill (1.9,0.2) circle (0.06cm);
\fill (2.1,0.2) circle (0.06cm);
 \node at (-0.3,-0.3) {\scriptsize $\Sigma^+$};
  \node at (4.2,-0.3) {\scriptsize $P$};
  \node at (2,0.8) {\scriptsize $H_W$};
% quarks
 \node at (1.0,0.38) {\tiny $s$};
  \node at (3.0,0.38) {\tiny $d$};
  \node at (3.0,-0.15) {\tiny $u$};
 \node at (1.0,-0.15) {\tiny $u$};
  \node at (2.0,-0.68) {\tiny $u$};
    \node at (2.0,-1.5) {\small C$_{su}$};
\end{scope}
%
%eye
%
\begin{scope}[xshift=11cm]
  \draw[bend right=35] (2,1.18) to (0,0);
  \draw[bend right=35] (4,0) to (2,1.18);
  \draw (2,0.62) circle (0.4cm);
%  \draw[thick] (0,0) to (4,0); 
 \draw (4,0) to (0,0);
 \draw[bend right=45] (0,0) to (4,0);
\fill (0,0) circle (0.06cm);
\fill (4,0) circle (0.06cm);
%Weak Hamiltonian
\fill (2,1.02) circle (0.06cm);
\fill (2,1.18) circle (0.06cm);
 \node at (-0.3,-0.3) {\scriptsize $\Sigma^+$};
  \node at (4.2,-0.3) {\scriptsize $P$};
  \node at (2,1.6) {\scriptsize $H_W$};
%quarks
 \node at (2.0,-0.17) {\tiny $u$};
  \node at (2.0,-0.68) {\tiny $u$};
  \node at (2.0,0.43) {\tiny $u,\!c$};
   \node at (0.8,0.70) {\tiny $s$};
  \node at (3.2,0.70) {\tiny $d$};
\node at (2.0,-1.5) {\small E};
\end{scope}
%
%saucer
%
\begin{scope}[xshift=16.5cm]
   \draw (2,1.1) circle (0.4cm);
  \fill[white] (1.9,0.8) -- (2.1,0.8) -- (2.1,0.6) -- (1.9,0.6) --  (1.9,0.8);
 \draw[bend right=25] (1.9,0.7) to (0,0);
  \draw[bend right=25] (4,0) to (2.1,0.7);
%    \draw[thick][bend left=35] (1.9,0.2) to (0,0);
%   \draw[thick][bend left=35] (4,0) to (2.1,0.2);
 \draw (4,0) to (0,0);
 \draw[bend right=45] (0,0) to (4,0);
\fill (0,0) circle (0.06cm);
\fill (4,0) circle (0.06cm);
%Weak Hamiltonian
\fill (1.9,0.7) circle (0.06cm);
\fill (2.1,0.7) circle (0.06cm);
 \node at (-0.3,-0.3) {\scriptsize $\Sigma^+$};
  \node at (4.2,-0.3) {\scriptsize $P$};
  \node at (2,0.4) {\scriptsize $H_W$};
%
%quarks
 \node at (2.0,-0.17) {\tiny $u$};
  \node at (2.0,-0.68) {\tiny $u$};
  \node at (2.0,1.22) {\tiny $u,\!c$};
   \node at (0.7,0.68) {\tiny $s$};
  \node at (3.3,0.70) {\tiny $d$};
\node at (2.0,-1.5) {\small S};
\end{scope}

  \end{tikzpicture}
  
\caption{Four diagram typologies for the $s \to d$ weak Hamiltonian in the $\Sigma^+ \to p$ transition. These are the Connected ($C_{sd}$ and $C_{su}$), Eye (E) and Saucer (S) diagrams.}
\label{Fig:S2N_Hw_3pt}
\end{figure}